\documentclass[journal]{IEEEtran}
\makeatletter

\usepackage{blindtext}
\usepackage[T1]{fontenc}
\usepackage{textcomp}
\usepackage{amsmath}
\usepackage{amssymb}
\usepackage{bm}
\usepackage{mdwmath}
\usepackage{cases}
\usepackage{graphicx}
\graphicspath{{Figure/PDF/}{Biography/PDF/}}
\usepackage[dvipsnames]{xcolor}
\definecolor{myblue}{rgb}{0,0.4980,1} 
\definecolor{myred}{rgb}{0.8706,0.1608,0.0627} 
\usepackage[caption=false,font=footnotesize,subrefformat=parens]{subfig}
\usepackage[nosort,noadjust]{cite}
\usepackage{url}
\usepackage{tabularray}
\UseTblrLibrary{counter}

\usepackage{hyperref}
\newcommand{\colorhypersetup}{\@ifpackageloaded{hyperref}{\hypersetup{%
	bookmarksopen=true,%
	bookmarksnumbered=true,%
	pdfpagemode={UseOutlines},
	pdfstartview={FitH},%
	colorlinks=true,%
	linkcolor={myred},%
	citecolor={orange}
}}{\empty}}
\newcommand{\blackhypersetup}{\@ifpackageloaded{hyperref}{\hypersetup{%
	bookmarksopen=true,%
	bookmarksnumbered=true,%
	pdfpagemode={UseOutlines},
	pdfstartview={FitH},%
	colorlinks=true,%
	allcolors={black}
}}{\empty}}
\blackhypersetup

\usepackage{mfirstuc-english}
\MFUhyphentrue
\usepackage[version2,description,IEEEtran]{eacro}
\acsetup{%
    page-style=paren,%
    hyperref=false%
}
\DeclareAcronym{SNR}{
	short = SNR,
	long = signal-to-noise ratio}
\DeclareAcronym{mmWave}{
	short = mmWave,
	long = millimeter wave}
\DeclareAcronym{5G}{
	short = 5G,
	long = 5th generation}
\DeclareAcronym{MIMO}{
	short = MIMO,
	long = multiple-input and multiple-output}
\DeclareAcronym{BS}{
	short = BS,
	long = base station}
\DeclareAcronym{LOS}{
	short = LoS,
	long = line of sight}
\DeclareAcronym{CSI}{
	short = CSI,
	long = channel state information}
\DeclareAcronym{UE}{
	short = UE,
	long = user equipment}
\DeclareAcronym{ML}{
	short = ML,
	long = machine learning}
\DeclareAcronym{CV}{
	short = CV,
	long = computer vision}
\DeclareAcronym{UAV}{
	short = UAV,
	long = unmanned aerial vehicle}
\DeclareAcronym{YOLO}{
	short = YOLO,
	long = You Only Look Once}
\DeclareAcronym{SSD}{
	short = SSD,
	long = single shot detector}
\DeclareAcronym{RCNN}{
	short = R-CNN,
	long = region based convolutional neural network}
\DeclareAcronym{2D}{
	short = 2D,
	long = two-dimension}
\DeclareAcronym{3D}{
	short = 3D,
	long = three-dimension}
\DeclareAcronym{RGB}{
	short = RGB,
	long = {red, green and blue}}
\DeclareAcronym{ROI}{
	short = RoI,
	long = region of interest}
\DeclareAcronym{MAP}{
	short = mAP,
	long = mean average precision}
\DeclareAcronym{AP}{
	short = AP,
	long = average precision}
\DeclareAcronym{FPS}{
	short = FPS,
	long = frames per second}
\DeclareAcronym{PR}{
	short = PR,
	long = precision-recall}
\DeclareAcronym{NLOS}{
	short = NLoS,
	long = non-line of sight}
\DeclareAcronym{FDD}{
	short = FDD,
	long = frequency division duplex}
\DeclareAcronym{RIS}{
    short = RIS,
    long = reconfigurable intelligent surface}
\DeclareAcronym{RF}{
    short = RF,
    long = radio frequency}
\DeclareAcronym{DFRC}{
    short = DFRC,
    long = dual-function radar communication}
\DeclareAcronym{ARPN}{
    short = ARPN,
    long = attention-based region proposal network}
\DeclareAcronym{RPN}{
    short = RPN,
    long = region proposal network}
\DeclareAcronym{ULA}{
    short = ULA,
    long = uniform linear array}
\DeclareAcronym{CDF}{
    short = CDF,
    long = cumulative distribution function}

\usepackage{algpseudocode}
\newcounter{MYalgorithmic}

{%
	\vspace*{3pt}
	\hrule height 1pt}
\newcounter{MYitem}[MYalgorithmic]

\newcommand{\MYlabel}[1]{\def\@currentlabel{\theALG@line}\label{#1}}

\newcommand{\upperroman}[1]{\uppercase\expandafter{\romannumeral#1}}

\newcommand{\myvec}[1]{\bm{\mathrm{#1}}}

\newcommand{\myunit}[1]{%
	\ifmmode
		\mathrm{#1}
	\else
		$ \mathrm{#1} $
	\fi}
\newcommand{\murm}{%
	\ifmmode
		\text{\textmu}
	\else
		\textmu
	\fi}

\newcommand{\MYnewpage}{%
	\ifCLASSOPTIONonecolumn
		\ifCLASSOPTIONjournal
			\typeout{The onecolumn journal mode.}
			\newpage
		\fi
	\fi}

\newlength{\mysinglefigwidth}
\newlength{\mymultifigwidth}
\ifCLASSOPTIONonecolumn
	\AtBeginDocument{\setlength{\mysinglefigwidth}{0.7\linewidth}}
\else
	\AtBeginDocument{\setlength{\mysinglefigwidth}{\linewidth}}
\fi
\ifCLASSOPTIONonecolumn
	\AtBeginDocument{\setlength{\mymultifigwidth}{0.5\linewidth}}
\else
	\AtBeginDocument{\setlength{\mymultifigwidth}{\linewidth}}
\fi

\makeatother
\begin{document}
\ifCLASSOPTIONonecolumn
	\typeout{The onecolumn mode.}
	\title{Vision-Assisted mmWave Beam Management for Next-Generation Wireless Systems: Concepts, Solutions and Open Challenges}
	\author{Kan~Zheng,~\IEEEmembership{Senior~Member,~IEEE}, Haojun~Yang,~\IEEEmembership{Member,~IEEE}, Ziqiang~Ying, Pengshuo~Wang, and~Lajos~Hanzo,~\IEEEmembership{Life~Fellow,~IEEE}
	}
\else
	\typeout{The twocolumn mode.}
	\title{Vision-Assisted mmWave Beam Management for Next-Generation Wireless Systems: Concepts, Solutions and Open Challenges}
	\author{Kan~Zheng,~\IEEEmembership{Senior~Member,~IEEE}, Haojun~Yang,~\IEEEmembership{Member,~IEEE}, Ziqiang~Ying, Pengshuo~Wang, and~Lajos~Hanzo,~\IEEEmembership{Life~Fellow,~IEEE}
	}
\fi

\ifCLASSOPTIONonecolumn
	\typeout{The onecolumn mode.}
\else
	\typeout{The twocolumn mode.}
\fi

\maketitle

\ifCLASSOPTIONonecolumn
	\typeout{The onecolumn mode.}
	\vspace*{-50pt}
\else
	\typeout{The twocolumn mode.}
\fi
\begin{abstract}
Beamforming techniques have been widely used in the millimeter wave (mmWave) bands to mitigate the path loss of mmWave radio links as the narrow straight beams by directionally concentrating the signal energy. However, traditional mmWave beam management algorithms usually require excessive channel state information overhead, leading to extremely high computational and communication costs. This hinders the widespread deployment of mmWave communications. By contrast, the revolutionary vision-assisted beam management system concept employed at base stations (BSs) can select the optimal beam for the target user equipment (UE) based on its location information determined by machine learning (ML) algorithms applied to visual data, without requiring channel information. In this paper, we present a comprehensive framework for a vision-assisted mmWave beam management system, its typical deployment scenarios as well as the specifics of the framework. Then, some of the challenges faced by this system and their efficient solutions are discussed from the perspective of ML. Next, a new simulation platform is conceived to provide both visual and wireless data for model validation and performance evaluation. Our simulation results indicate that the vision-assisted beam management is indeed attractive for next-generation wireless systems.
\end{abstract}

\ifCLASSOPTIONonecolumn
	\typeout{The onecolumn mode.}
	\vspace*{-10pt}
\else
	\typeout{The twocolumn mode.}
\fi

\IEEEpeerreviewmaketitle

\MYnewpage


\section{Introduction}
\label{sec:Introduction}

\acresetall


\IEEEPARstart{B}{eamforming}-aided directional transmission plays a critical role in improving the spatial spectrum efficiency. Due to the expected wide deployment of \ac{mmWave} communications, beamforming techniques are receiving much attention in the context of \ac{MIMO} systems designed for the \ac{mmWave} frequency bands~\cite{Bjornson2019}. However, given a large number of antennas, tracking the movement of multiple concurrent \acp{UE} dramatically increases the complexity, overhead and latency of signal processing at the \acfp{BS} using \ac{mmWave} massive \ac{MIMO} schemes~\cite{Liu2020}. These problems may be exacerbated for a high number of antennas even in \acf{LOS} channels. In order to overcome these challenges, \ac{CV}-aided \ac{ML} algorithms may be harnessed as promising solutions for beamforming. Motivated by the spatial sparsity of \ac{mmWave} wireless channels exhibiting predominant \ac{LOS} characteristics, \ac{mmWave} beams pointing to the target \acp{UE} can be efficiently selected and adapted according to the location information of \acp{UE} derived by \ac{ML} algorithms~\cite{Charan2021}.

In order to implement a vision-assisted beam management system, several technical challenges have to be overcome. The traditional vision-based object tracking algorithms, such as the sparse representation and correlation filtering, have difficulty in accurately locating high-mobility \acp{UE} in real-time~\cite{Druzhkov2016}. Furthermore, in complex environments, tracking multiple \acp{UE} in the face of blockage and uneven light, the localization accuracy of \acp{UE}  tends to degrade significantly. As a result, the traditional \ac{CV}-related \ac{ML} algorithms cannot satisfy the high location accuracy required by \ac{mmWave} communications. Finally, gathering abundant labelled data from real-world environments including both visual data and wireless signals to train \ac{ML} models is still challenging.

\begin{figure*}[!t]
	\centering
	\includegraphics[width=0.7\linewidth]{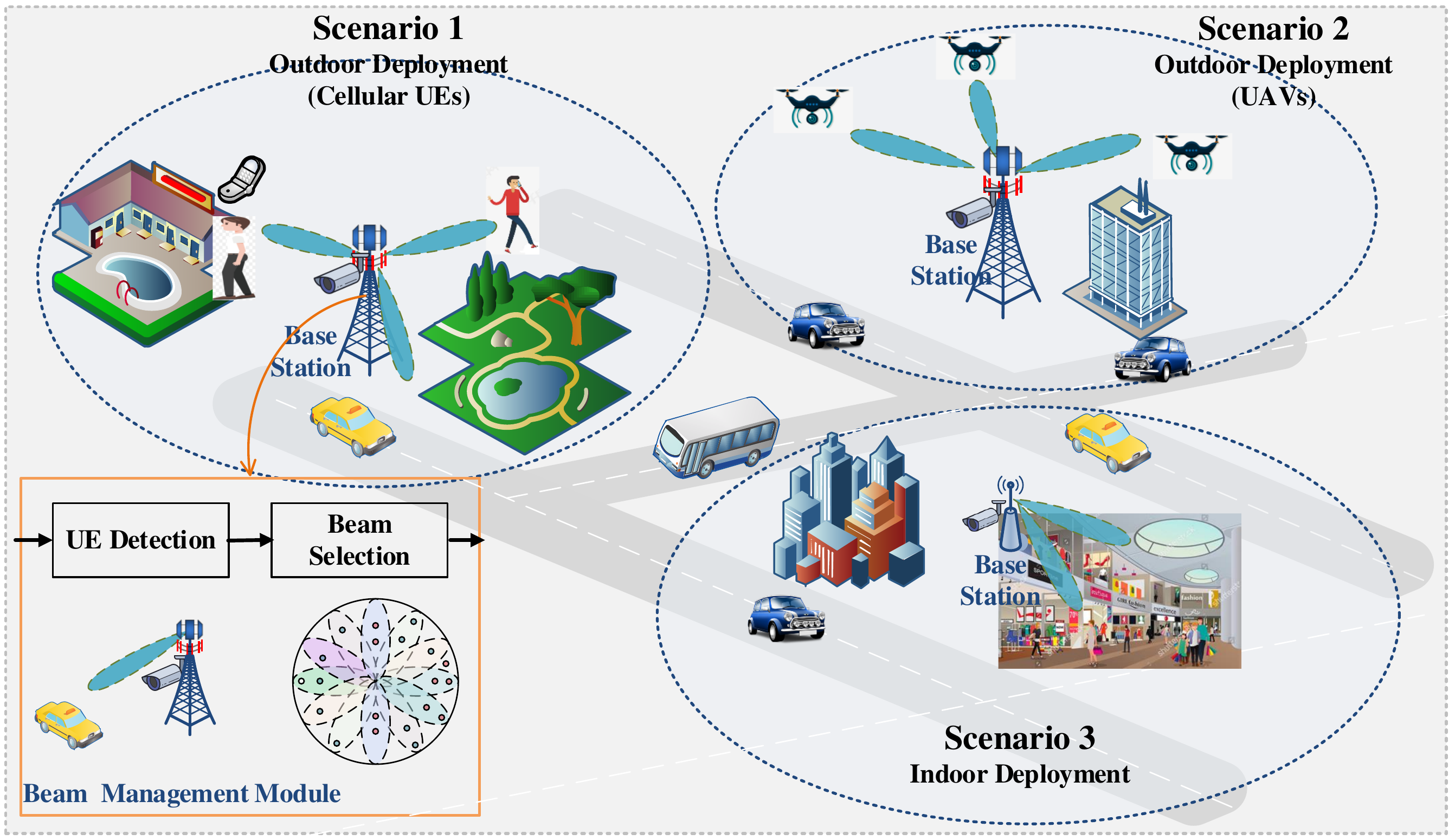}
	\caption{Illustration of vision-assisted beam management system.}
	\label{fig_system}
\end{figure*}

Nevertheless, vision-assisted beam management methods that predict the optimal \ac{mmWave} beams have been investigated in the last few years. A framework used for dataset generation was also proposed for cooperatively exploiting both visual and wireless data~\cite{ViWi}. However, these methods may still be plagued by a number of issues: 
\begin{itemize}
\item The robustness of the existing \ac{ML} models for vision-assisted beam management has to be improved~\cite{Tian2021}. When the classical image classification models designed for prediction are used for \ac{mmWave} link blockage prediction and beam prediction, the target accuracy cannot be always satisfied, for example due to the over-fitting issues. In this context, a preliminary study was conducted in~\cite{Ahn2022} by relying on a simple dataset, for investigating vision-assisted beam management in multi-user scenarios.

\item The scalability of the methods is not guaranteed in practical scenarios. For example, the existing methods do not obey the so-called modular design principles, making them difficult to upgrade flexibly or to modify them simply when for example a crucial parameter, like the size of the beamforming codebook, changes.

\item The implementation issues pertaining to the complexity of computation and overhead costs have to be addressed before the wide deployment of vision-assisted beam management becomes a reality. For example, an exploratory strategy was proposed for reducing the overhead associated with beam selection, where information from localization and vision sensors is integrated~\cite{Reus2021}.
\end{itemize}

Indeed, there is a paucity of literature addressing the associated challenges of vision-assisted \ac{mmWave} beam management techniques. The scope of this article is thus to study the interplay of \ac{ML}-based \acl{CV} and beam management in \ac{mmWave} systems. Specifically, the main contributions of this article can be summarized as follows.
\begin{enumerate}
\item We first present a comprehensive framework for a vision-assisted \ac{mmWave} beam management system, including the typical deployment scenarios as well as a pair of major concerns, namely the user equipment detection and beam selection.

\item Then, three main technical challenges and their efficient solutions are discussed from the perspective of \ac{ML}. In particular, we study the salient issues of lightweight compression, the deleterious effects of inadequately labeled data, as well as the associated robustness aspects.

\item Next, we portray the development of our own simulation platform to provide both visual and wireless data for model validation and performance evaluation. This unified platform is universally applicable in terms of producing data for those scenarios where the wireless characteristics vary tremendously. Our simulation results also show that vision-assisted beam management is indeed attractive for next-generation wireless systems.

\item Lastly, the related open topics are discussed from a practical perspective in order to guide future research.
\end{enumerate}

The article is organized as follows. A detailed description of vision-assisted \ac{mmWave} beam management systems is provided in Section~\ref{sec:framwork}. Next, we discuss some \ac{ML}-related challenges and solutions conceived for \ac{mmWave} beam management in Section~\ref{sec:challenge}. We then present our performance results and discuss some potential open topics. Finally, our conclusions are given in Section~\ref{sec:Conclusion}.

\section{Holistic Framework of Vision-Assisted Beam Management System}
\label{sec:framwork}

\subsection{Typical Deployment Scenarios of Vision-Assisted Beam Management Systems}

The next-generation wireless systems are expected to operate in multiple bands, including the sub-$6~\myunit{GHz}$ and \ac{mmWave} bands. In general, the signal propagation of the sub-$6~\myunit{GHz}$ bands is more resilient to blockages, thereby the sub-$6~\myunit{GHz}$ bands are used for the services that require low or medium data rates. By contrast, as a benefit of their abundance of spectral resources, the \ac{mmWave} bands are expected to support multi-Gigabit services. In order to take full advantage of their benefits, a dual-band system in which the \ac{BS} and \acp{UE} use both the sub-$6~\myunit{GHz}$ and \ac{mmWave} transceivers is considered in this article. The vision-assisted beam management may be enabled only when the \ac{LOS} condition is met, which may also be combined with sub-$6~\myunit{GHz}$ systems~\cite{Charan2021}. The rich bandwidth potential of \ac{mmWave} communications can be used both for the backhauls and for the user access links under a variety of potential deployment scenarios.Thus, we mainly focus attention on those scenarios, where the vision-assisted beam management can be harmoniously integrated. Some of them are illustrated in Fig.~\ref{fig_system}, and are discussed in more detail below.

\subsubsection{Scenario 1 -- Outdoor Deployment (Cellular UEs)}

When the wireless channels under outdoor environments are spatially sparse, i.e., dominated by \ac{LOS} propagation, vision-assisted beam management can be indeed conveniently adopted at the \acp{BS}. Then, all the cellular \acp{UE} can be served by \acp{BS} on the \ac{mmWave} band.
 
\subsubsection{Scenario 2 -- Outdoor Deployment (UAVs)}

At the time of writing, \ac{mmWave} communications are widely used for \ac{UAV} communications. The camera deployed at \acp{BS} can also readily capture the video of the \ac{UAV} flying by without obstruction. Therefore, it is eminently suitable for vision-assisted beam management in this scenario.

\subsubsection{Scenario 3 -- Indoor Deployment}

In order to significantly increase the system capacity in high-density indoor environments, cameras installed at the \acp{BS} are capable of capturing images of nearby \acp{UE}. However, there may be lots of objects, which increases the recognition complexity of the \ac{CV} algorithms.

\subsection{Description of Vision-Assisted Beam Management System}

As shown in Fig.~\ref{fig_system}, a \ac{BS} equipped with a high-definition camera first captures the video scenes. Then, the \ac{ML}-based vision model embedded in the \ac{BS} is activated for localizing and tracking the target \acp{UE}. By collaboratively utilizing the image/video information, the beam management module finally selects the optimal beams for the target \acp{UE} among a pre-defined beam pattern codebook.

The vision-assisted beam management is generally divided into two steps, namely the \ac{UE} detection, and the ensuing beam selection for the target \acp{UE}. More explicitly, the former determines whether any target \acp{UE} exist in the view of the camera, while the latter is responsible for providing both the location information and the optimal beams for the target \acp{UE}.

\subsubsection{\acs*{UE} Detection}

In the traditional \ac{ML}-based object detection models, each frame of the video stream is processed to generate the object locations as the output, which is usually time-consuming. However, the target \acp{UE} are not captured by the camera all the time, and they are not always in their active communication status. To this end, it is necessary to detect the existence of active \acp{UE} before beam selection. In the proposed framework, a \ac{ML}-based binary classification model can be used for determining whether active \acp{UE} exist at the current moment. For the sake of illustration, the active state of the $ k $-th \ac{UE} is defined as ``1'', while  the inactive state as ``0'' . Then, based on the captured image, the active/passive state of the $ k $-th \ac{UE} can be predicted by
\begin{align}
S_k = F_{\mathcal{P}_1}\left( \mathsf{Image}, k \right),
\end{align}
where $ F_{\mathcal{P}_1}(\cdot) $ is the \ac{ML}-based binary classification model that has to be investigated, and $ \mathcal{P}_1 $ is the parameter set of the model. Additionally, to strike an attractive tradeoff between the complexity and accuracy, the \ac{UE}-related information including the sub-6\myunit{GHz} channel state information and network signaling might be taken into account.

\subsubsection{Beam Selection for Target UEs}

The goal of beam selection is to find the optimal beam from the codebook for maximizing the \ac{SNR}. The traditional beam management schemes generally require the \ac{CSI} to be obtained by channel estimation, which requires substantial overhead. Instead, a vision-assisted method requiring no \ac{CSI} knowledge is conceived for solving the beam selection problem. Firstly, the position of target \acp{UE} can be determined by a \ac{ML}-based object detection model from the images captured by the camera. Then, the angles of the target \acp{UE} are estimated by exploiting the location information. Finally, the optimal beam index is selected by maximizing the \ac{SNR}, albeit other metrics may also be used. The complete procedure is as follows.

\paragraph{\textbf{Object Detection}}
Given the presence of some target \acp{UE}, each frame of the video stream can be processed to locate the target \acp{UE} by \ac{ML}-based object detectors. In general, the family of \ac{ML}-based object detectors may be divided into two types, namely the single-stage detectors, such as the so-called \acl*{YOLO} type models~\cite{Yolov3}, and the two-stage detectors, such as \ac{RCNN} related models. The two-stage detector first adopts a `region proposal network' for generating the \ac{ROI}, and then utilizes classification models for determining the category of region. In contrast to the two-stage detector, the single-stage one directly predicts the category of each feature map without first generating \ac{ROI}. Hence, the two-stage detector typically attains higher detection accuracy, while the single-stage detector has higher detection speed. In our proposed framework, either of them may be chosen flexibly according to the specific requirements of different application scenarios.

\paragraph{\textbf{Angle Prediction}}
For the beam management, the angle information of the \acp{UE}' physical location within the geographical coverage of the \ac{BS} is required for selecting the optimal beam in terms of the real physical \textit{world coordinate}. However, the outputs of \ac{ML}-based object detection models are the \acp{UE}' location in the image captured by the camera, i.e., the location in the \textit{pixel coordinate}. Thus, it is paramount to establish the mapping relationship between these two locations in the cases of vision-assisted beam management applications.

\paragraph{\textbf{Beam Selection}}
Given the predicted angle, the beam selection can provide the index of the optimal beam. Let $ \myvec{w}_k $ denote the beamforming vector of the $ k $-th \ac{UE}. Then the optimal beam can be predicted as follows: 
\begin{align}
\myvec{w}_k = G_{\mathcal{P}_2}\left( \mathsf{Angle}, \mathsf{Codebook}, k \right),
\end{align}
where $ G_{\mathcal{P}_2}(\cdot) $ and $ \mathcal{P}_2 $ are the prediction model and parameter set, respectively. For instance, upon considering the simple case of a uniform linear array in the \acs{2D} space, the codebook is composed of $ Q $ beams having an identical angular separation of $ \pi/Q $. Therefore, the task of the beam selection is simplified to estimating the range that the predicted angle falls into.

\section{Challenges for Vision-Assisted Beam Management System}
\label{sec:challenge}

\begin{figure*}[!t]
	\centering
	\includegraphics[width=0.7\linewidth]{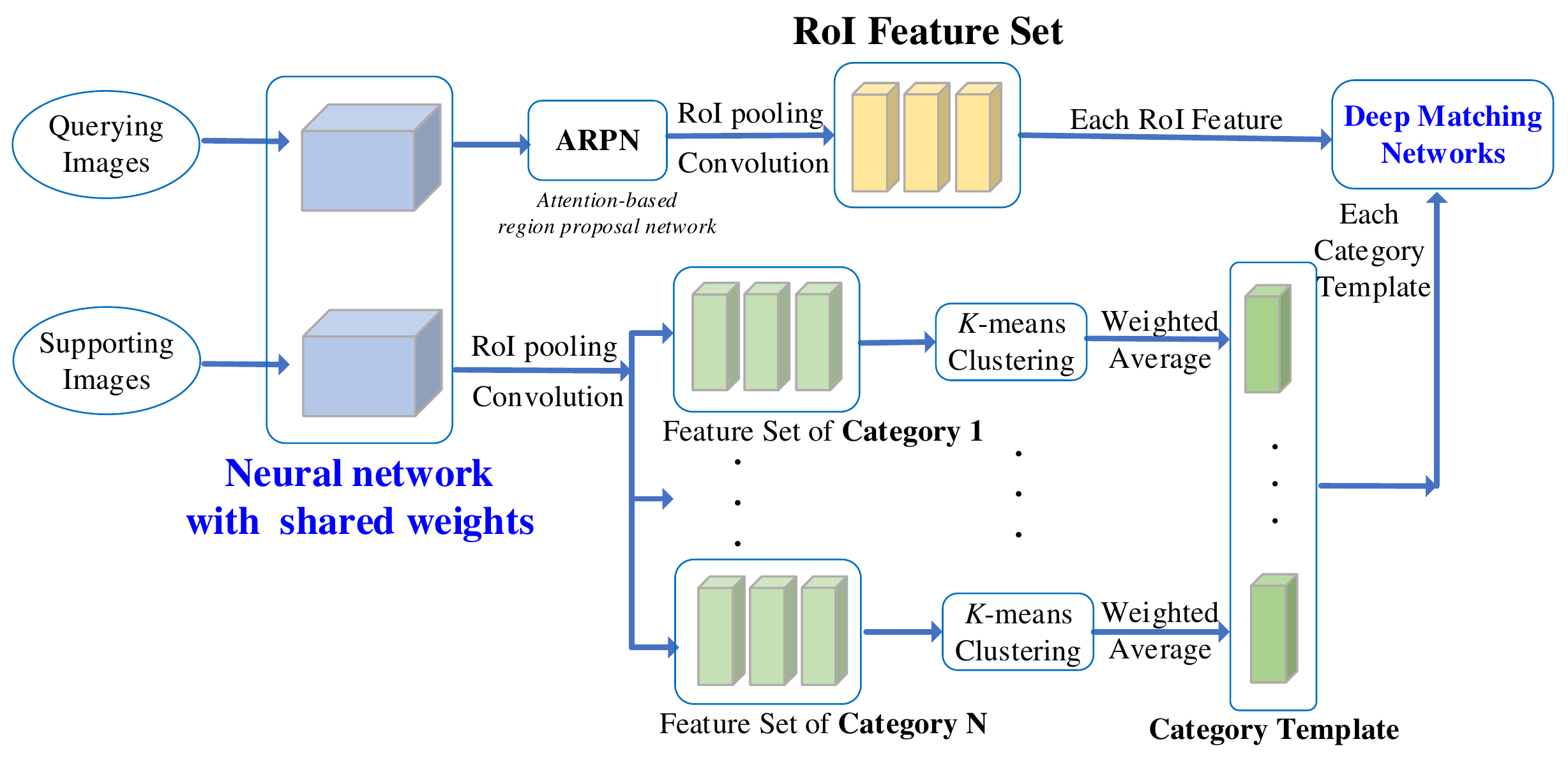}
	\caption{Illustration of $ N $-way $ K $-shot \acs*{ML} models based on metric learning for object detection.}
	\label{fig_few_shot}
\end{figure*}

\begin{figure*}[!t]
	\centering
	\includegraphics[width=0.7\linewidth]{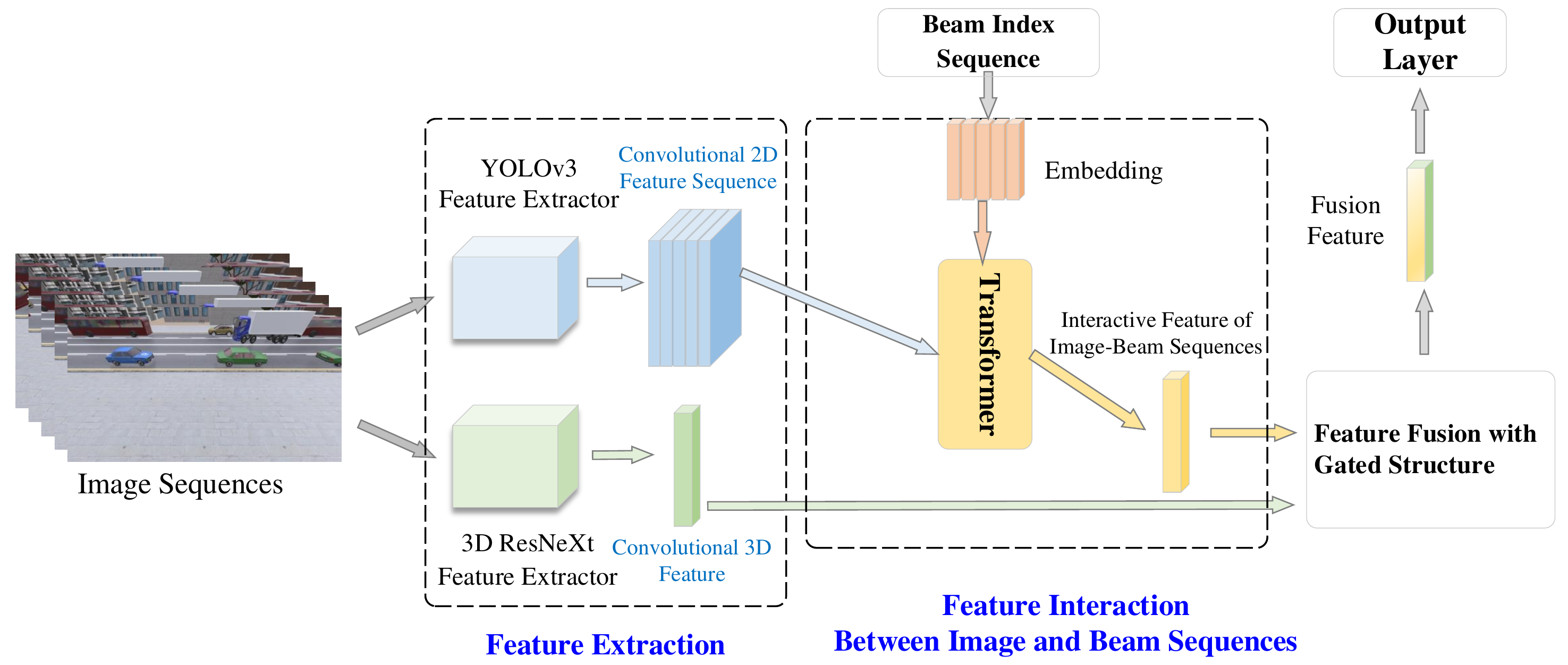}
	\caption{Illustration of a vision-assisted beam management system based on image sequences, where the structures of the \ac{YOLO}v3 and Transformer are the classical ones, and the 3D ResNeXt-101 is shown in~\cite{Hara2018}.}
	\label{fig_robust}
\end{figure*}

\subsection{Lightweight Compression for Prediction Model}

The limited computing and storage capabilities of embedded systems make the real-time implementation of the \ac{ML}-based models in \ac{mmWave} communication systems challenging. Again, the \ac{YOLO} object detector is applicable to localize the target \ac{UE}. Even though \ac{YOLO} is faster than other detectors, it still contains too many convolutional layers. For example, the backbone network in Version 3 of \ac{YOLO} (called `\ac{YOLO}v3'~\cite{Yolov3}) comprises $ 53 $ convolutional layers, and the channels in each of these convolutional layers are typically quite large, namely up to $ 1024 $ channels. Hence, the model size and computation complexity of \ac{YOLO} become the barriers to its time-critical applications such as \ac{mmWave} beam management. Therefore, it is essential to compress the model volume for increasing its prediction speed, while guaranteeing its accuracy.

One of the most common methods of model compression is network pruning~\cite{loffe2015}. In this method, sophisticated rules can be applied to neural networks so that the relatively insignificant weights or branches are removed, thereby reducing the number of model parameters and increasing the inference speed. According to the granularity of pruning objects, the typical network pruning schemes can be divided into weight pruning and structured pruning techniques. The former compresses those relatively insignificant weights in the networks. This technique has a high degree of flexibility, but modest inference speed acceleration. For the latter one, the coarser-grained convolution kernels, channels, and layers might be removed, resulting in both a higher compression ratio and faster inference speed.

Since \ac{YOLO} contains a large number of convolutional layers and hundreds or thousands of channels, a structured pruning method is preferred for obtaining a satisfactory compression effect. In particular, \ac{YOLO} can be pruned at both \textit{the channel and layer levels}. Channel pruning provides the compression of the model width, while the layer pruning reduces the depth of the models. With the help of network pruning, the volume of \ac{YOLO} model can be substantially reduced, hence its prediction speed is significantly improved.

\subsection{Efficiency Improvement of Prediction Model Having Inadequately Labeled Data}

Due to a large number of parameters in the \ac{ML}-based models, a dataset having a huge number of labeled data is required for training the models. However, there might be insufficient labeled data to fully fit the \ac{ML} models in practical vision-assisted \ac{mmWave} communication systems. First of all, gathering visual data (such as \acs{RGB} images) and wireless data (such as channel responses) requires completely different equipment and devices. Furthermore, realistic physical test scenarios have to be constructed, relying on practical equipment placing and data synchronization. Additionally, a long test period is needed in order to collect enough data. As a result, the data collection process itself is complex and time-consuming. Finally, the visual datasets collected have to label the bounding boxes for all \acp{UE} in the images. Thus, for practical applications, another challenge is how to achieve excellent prediction accuracy in the beam management module, when the dataset is small or moderate.

The output layer, which is used to map the feature vector to the required classification space, is typically a fully connected layer or a $ 1\times1 $-convolutional layer in both the single-stage and two-stage object detectors. In general, the output layer parameters of object detectors are randomly initialized and iteratively updated thereafter, when a new dataset is adopted in the training. However, there are also a number of other parameters in the models that have to be fine-tuned. As a result, having inadequate data may cause over-fitting during the learning process, gravely affecting the localization of objects. Localization performance degradation  may lead to the spurious angle prediction for beam selection algorithms.

Therefore, in order to improve the modelling process in the face of inadequately labeled data, we propose to use a \ac{ML} scheme relying on the \textbf{\textit{`metric learning}}\footnote{In general, many approaches in \ac{ML} require a measure of distance among data points. Typically, with the aid of priori domain knowledge, some standard distance metrics are adopted, such as Euclidean, Cosine, etc. Nevertheless, it is difficult to design metrics that are well-suited to the particular data and task of interest. Therefore, \textbf{\textit{`metric learning'}} technique is investigated to automatically construct task-specific distance metrics from weakly supervised data, which is more beneficial for the case of inadequately labeled data.}\textbf{\textit{'}} technique of~\cite{Fan2020} for accurately localizing the \acp{UE}. Fig.~\ref{fig_few_shot} presents a $ N $-way $ K $-shot \ac{ML} scheme conceived for object detection based on metric learning. In this scheme, a \ac{ROI} set is generated for the querying images based on region proposal networks. For the supporting images, the features of all categories are generated according to the labeled frames. Our proposed scheme calculates the similarity between each predicted \ac{ROI} feature and the corresponding feature template. Theoretically, the higher the similarity score, the higher the prediction accuracy becomes for the bounding box associated with the \ac{ROI}.

\subsection{Robustness and Applicability for Prediction Model}

Only adopting low-complexity single frame image based methods cannot cope well with multi-user scenarios, especially in cluttered environments. In the case of completely invisible \acp{UE}, the \acp{BS} cannot identify them, because they may be totally obscured when simply analyzing a single image frame at a time. Explicitly, single image contains only the location information and environmental information about the \acp{UE} seen at the time, but cannot provide extra information concerning the movement of the target \acp{UE} or the changing camera-view of the surrounding environment. Hence the single-frame  processing loses sight of the spatial and temporal correlation of moving objects. Therefore, how to exploit the image sequences in the video data to improve the performance of the object detectors and beam management remains a challenge.

To enhance the robustness and applicability, the vision-assisted beam management may process the image sequences for a total of $ N $ consecutive video frames, i.e., not only the current frame but also those from $ N-1 $ previous frames. Compared to the schemes based on the current individual video frame, the improved schemes using a sequence of video frames can capture both the spatial coherence of each video frame and its temporal inter-frame correlation.

Fig.~\ref{fig_robust} shows the overall framework of a vision-assisted beam management system based on image sequences~\cite{Yao2020}. The framework primarily consists of three main steps. In the first step, we extract specific features of the image sequences. In the lower branch of Fig.~\ref{fig_robust}, the \acs{3D} convolution is applied for extracting the features containing both spatial and temporal contents. In the upper branch of Fig.~\ref{fig_robust}, the \acs{2D} convolution is used for processing each image separately. Then, the interactions among the image-beam sequence features take place. The Transformer scheme of Fig.~\ref{fig_robust} having several encoder layers is used for interactive sequence modeling of the features in the beam index sequence and the image sequence features obtained by \acs{2D} convolution. The final step is to design a suitable output layer according to the specific beam management tasks so as to select the optimal beam.

\section{Simulation Methodology and Evaluation}

\begin{figure*}[!t]
	\centering
	\includegraphics[width=0.6\linewidth]{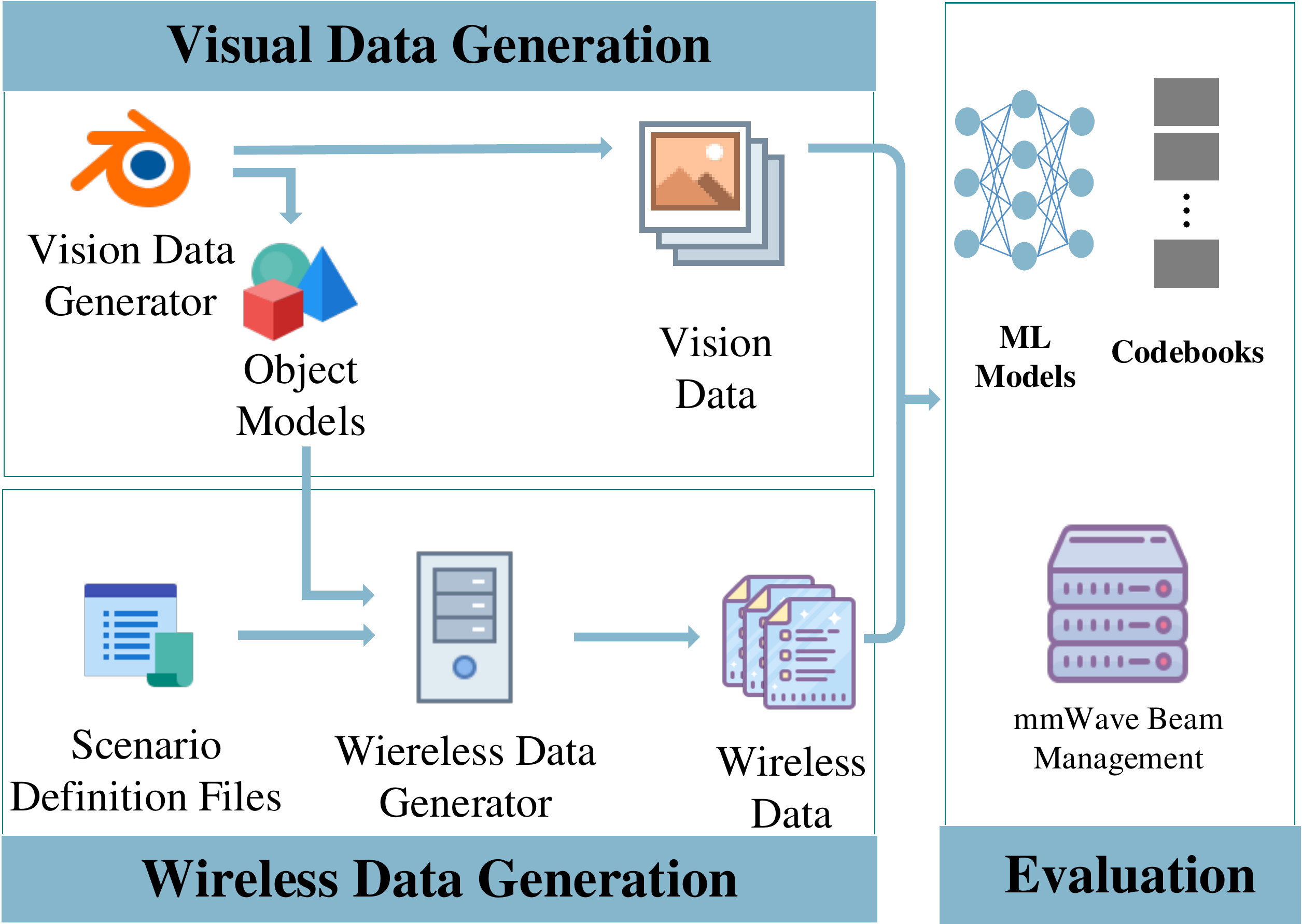}
	\caption{Simulation platform for vision-assisted beam management system.}
	\label{fig_sim}
\end{figure*}

At the current state-of-the-art, it is quite difficult to collect and label both the visual and wireless data in real-time. Hence, we resort to simulations for generating labeled data for training and testing. Fig.~\ref{fig_sim} shows our simulation platform conceived for vision-assisted \ac{mmWave} beam management. As this stage, only an outdoor scenario is used for validating the models in the platform. An open source framework is proposed to speed up the implementation of other scenarios. As a result, it is advantageous to revise the details of the scenario when generating wireless data, such as the number and orientation of rays, channels, user positions, etc. Furthermore, diverse \acp{UE} are involved, as well as other entities, such as trees, bushes, sidewalks, benches and buildings. Specifically, our own-developed and open-source platform is based on MATLAB software and only requires a text file for defining a scenario. During the phase of initialization, a series of visual and wireless sequences are created for modelling real-world physical environments. To create visual and wireless datasets, all sequences are respectively processed by the animation modeling software\footnote{Normally, the animation modeling software is designed for creating complex \acs{3D} objects, rendering them to images, and making animation from frames.} and the ray-tracing software in the second step. Finally, the datasets can be used for evaluating and validating the performance of \ac{ML}-based models for beam management.

\subsection{Initialization}

In the initialization phase, the types and attributes of entities are described in intricate detail. Using unified definitions is an efficient and compatible way of ensuring the appropriate relationship between the visual and wireless data generation. In particular, the scenario definition includes the system parameters, antenna arrays, \acp{BS}, reflectors, and mobile users. Each of them contains the following information, i.e.,
\begin{itemize}
    \item \textbf{System parameters:} This includes parameters used to describe how the platform works. For example, the total number of frames for simulations, the average number of video frames calculated per second in simulations, the maximum number of \ac{mmWave} reflections calculated in the ray tracing process, and the size of beam codebook, etc.
     \item \textbf{Antenna arrays:} The key parameters of the antennas such as the size of antenna arrays and the antenna spacing are defined.
    \item \textbf{\Aclp{BS}:} The location of \acp{BS}, the configuration of the camera deployed at \acp{BS}, the antenna arrays used by \acp{BS} and diverse other parameters are described in detail.
    \item \textbf{Reflectors:} The position, shape and material of reflectors are given.
    \item \textbf{\Aclp{UE}:} Similar to the reflectors, we have to define the parameters of \acp{UE} such as the location and appearance. Additionally, the \ac{UE} antenna arrays, the \ac{UE} motion trajectories and other parameters are specified as well.
\end{itemize}

It is noted that both the visual and wireless simulations require the aforementioned information. Based on the information defined in a given scenario, the visual and wireless simulation processes have to be synchronized so that the data generated tally correctly.

\subsection{Data Generation}

The data generation includes both visual and wireless data. Although different simulation environments and processes are used for generating these two types of data, there still exists a corresponding relationship between them for ensuring that the data produced conforms to the scenario definition. Here, we first introduce the process of generating both types of data, and then we describe how to 
synchronize and merge these data.

\subsubsection{Visual Data Generation}

The visual data is generated by some special animation modeling software, such as Blender~\cite{blender}, which facilitates the construction of \acs{3D} object models. Hence, the first step in generating the visual data is to create a \acs{3D} model of the reflectors and users by the animation modeling software. Then, we have to assign textures or materials to the objects in the scenario, in order to make them more realistic. Note that the material mentioned here differs from that in the scenario definition. The former only determines the visual effect of the generated image, while the latter determines the propagation of electromagnetic waves. Next, the cameras have to be deployed correctly at the \acp{BS}. The second step is to define the movement animation of users. In general, the animation consists of a sequence of images. There are some frames referred as the key frames, and the position and shape of the \acs{3D} model in the other frames can be determined by interpolation between a pair of consecutive key frames. In the scenario defined, the objects are regarded as rigid bodies, and each frame only contains the position change users. Finally, the  animation generated is the last step exported from the animation modeling software.

\subsubsection{Wireless Data Generation}

The wireless data is generated by the software that supports ray-tracing technology~\cite{raytracing}, e.g., MATLAB. Due to the challenge of generating complex \acs{3D} objects in MATLAB, the \acs{3D} model of the reflectors and users must be obtained by loading external data. Then, the transmitter and receiver are correctly positioned, i.e., co-located with either the \acp{BS} or the other \acp{UE} that might move at a given speed and in a certain direction. Next, we calculate the propagation-related information, such as the signal power, delay, angle of departure and angle of arrival, using ray-tracing technology. Likewise, according to the geometric channel model, the wireless channels in the current scenario are constructed using the above propagation information. Furthermore, the codebook indices corresponding to the optimal beam are calculated, which is crucial for the wireless data.

\subsubsection{Data Synchronization}


\begin{figure*}[!t]
\centering
\begin{tblr}{
    width = \linewidth,
    colspec = {X[1,c,m]X[1,c,m]},
    columns = {leftsep=0pt,rightsep=0pt},
}
    \subfloat[Illustration of an example of \acs*{mmWave} communication scenario.]{\includegraphics[scale = 0.49]{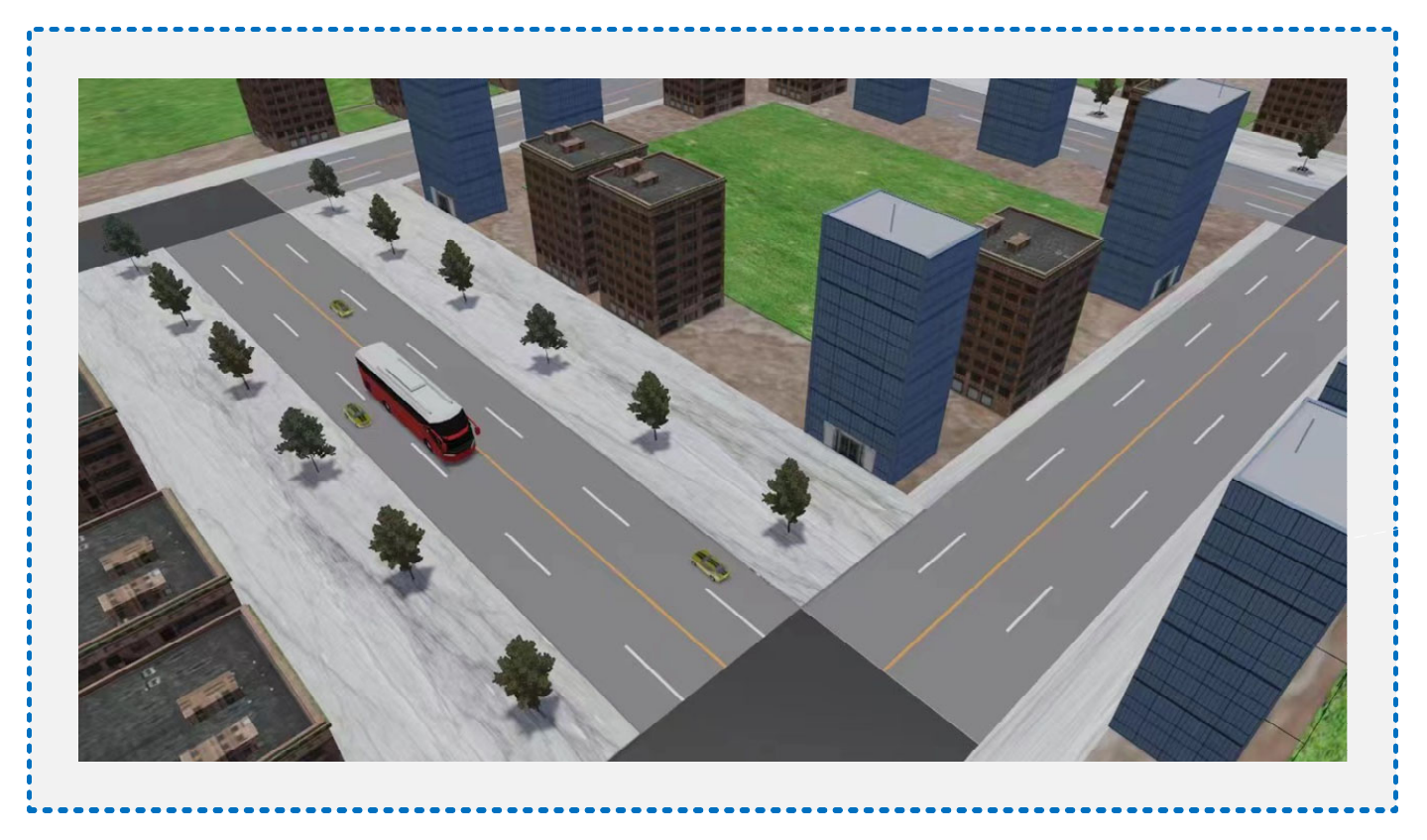}\label{fig_sub1}} & \subfloat[An example for single frame image with bounding box labelled by object detection.]{\includegraphics[scale = 0.49]{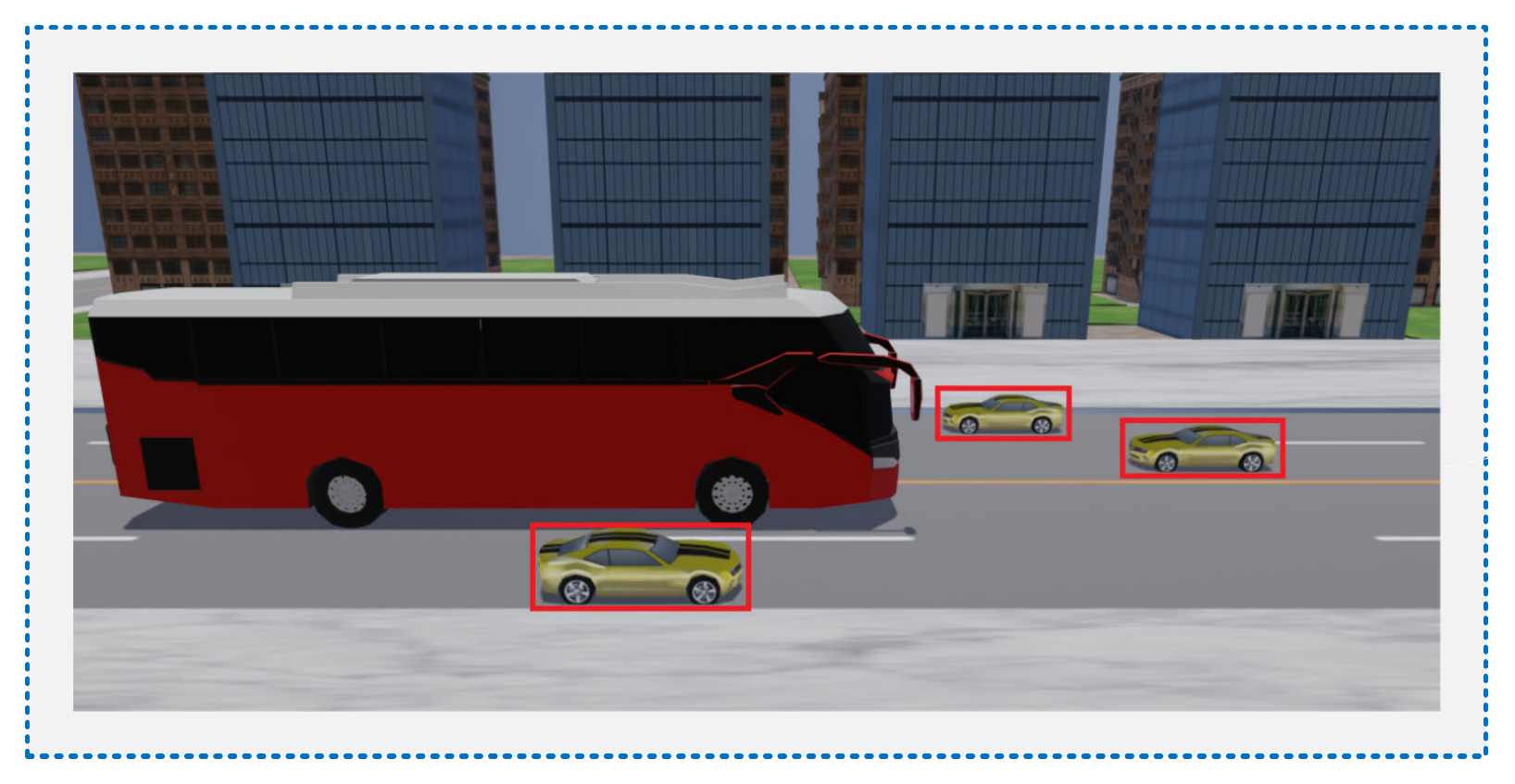}\label{fig_sub2}} \\
    \subfloat[An example for ray tracing at one frame.]{\includegraphics[width = \linewidth]{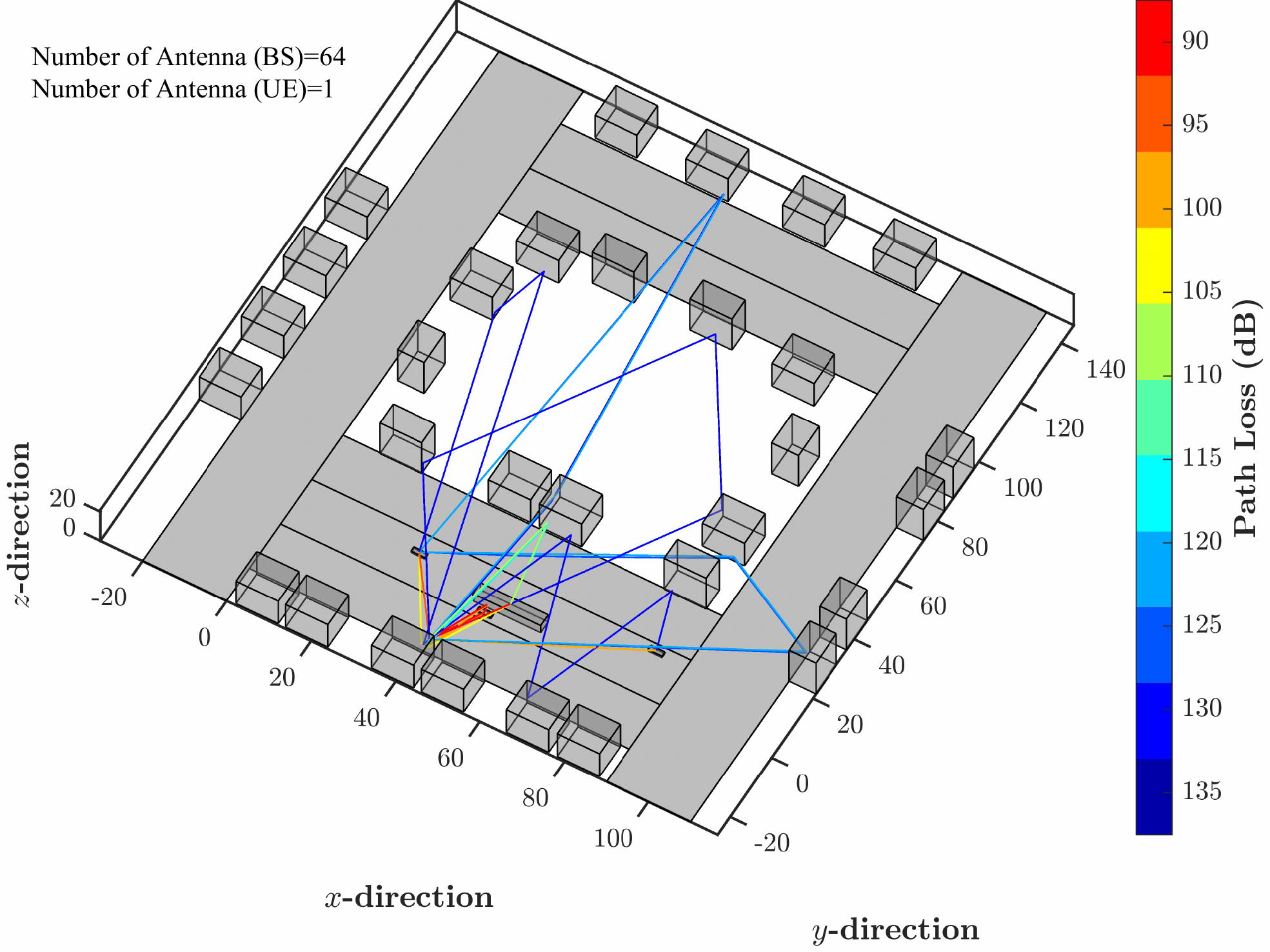}\label{fig_sub3}} & \subfloat[Performances of beam prediction.]{\includegraphics[width = \linewidth]{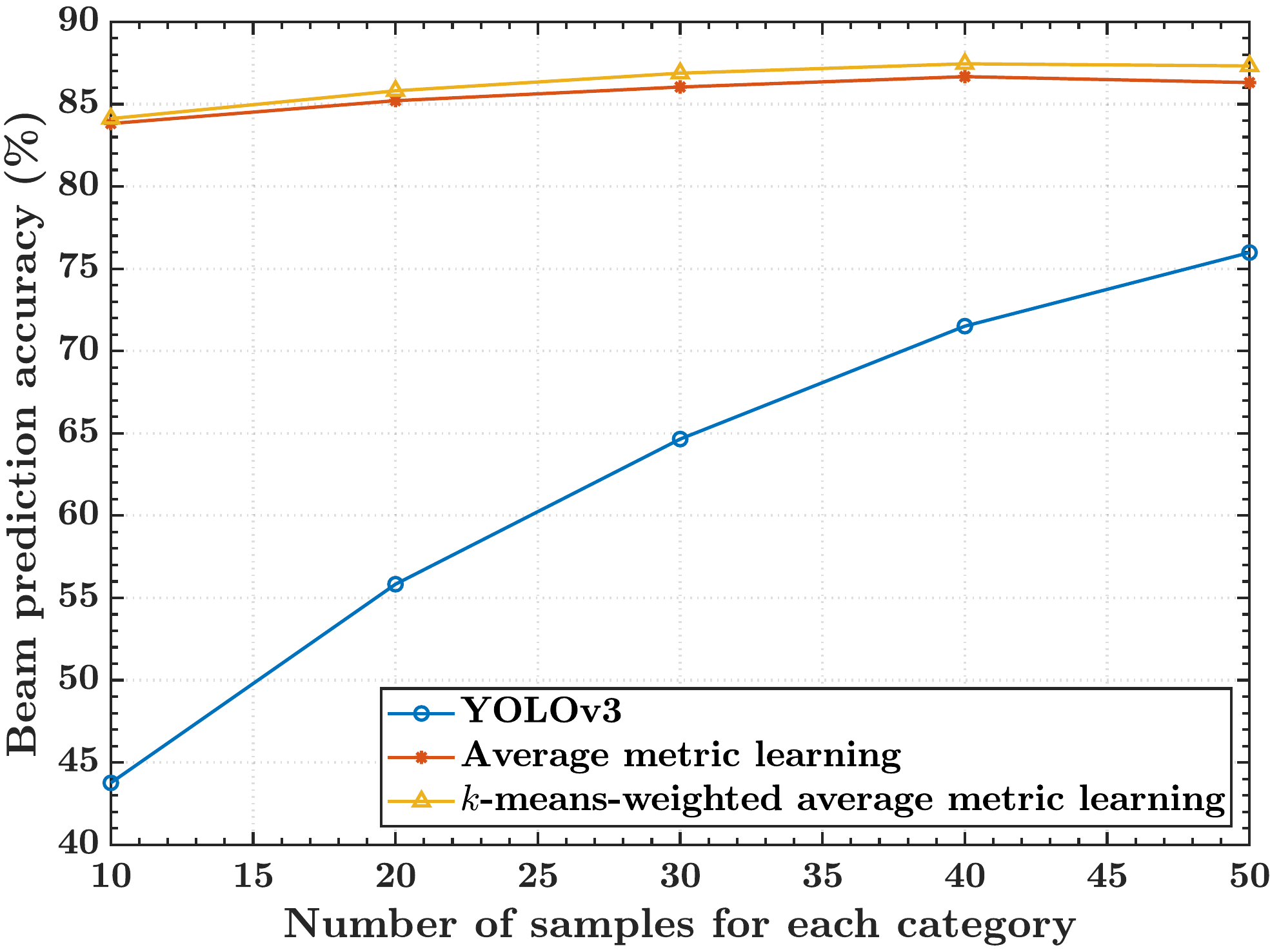}\label{fig_sub4}}
\end{tblr}
\caption{Performance evaluation results.}
\label{fig_test1}
\end{figure*}

As seen from the above data generation phase, both the object model and \ac{UE} motion should be consistent across the pair of generation processes. Specifically, the following methods are used in our platform.
\begin{itemize}
    \item Consistency of object model: With the aid of the animation modeling software, we create \acs{3D} models of the objects and export them in STL format\footnote{The STL format is a universal format for displaying \acs{3D} models, which is widely supported by related animation modeling software.}. Then, we import the required \acs{3D} model into MATLAB software with ray-tracing to make sure that the object models are consistent.
    \item Consistency of \ac{UE} movement: The concept of frames is introduced into the ray-tracing process. Hence, the \acp{UE} remain in the same positions for these two generation processes, which ensures that the \ac{UE} movements are consistent.
\end{itemize}

\subsection{Evaluation and Validation}

Fig.~\subref*{fig_sub1} shows an example of the \ac{mmWave} communication scenario in an urban environment, where a \ac{BS} communicates with three vehicular \acp{UE} in cars. Moreover, two types of buildings built from different materials are involved in this scenario. In general, the reflection properties of objects are strongly influenced by their materials. To accurately model real-world environments, we set the materials of the dark-coloured and light-colored buildings to concrete and brick, while the materials of all vehicles are assumed to be metal. We then characterise the performance of the platform based on this pre-defined scenario.


\renewcommand\TblrOverlap[1]{#1}
\begin{table*}[!t]
\centering
\caption{Simulation Results for \acs*{ML} models with Lightweight Compression}
\label{table_prune-perform-comp}
\begin{talltblr}[
    label = none,
    note{1} = {\Acl*{MAP}},
    note{2} = {\Acl{FPS}},
]{
    width = 0.9\linewidth,
    colspec = {X[5,l,m]X[2,c,m]X[3,c,m]X[3,c,m]X[3,c,m]X[1,c,m]},
    hlines,
    hline{3} = {1}{-}{},
    hline{3} = {2}{-}{},
    vline{2-6},
    cell{1}{1} = {r=2}{c,m},
    cell{1}{2} = {c=2}{c,m},
    cell{1}{4} = {c=3}{c,m},
    cell{3-5}{2-6} = {mode=dmath},
}
    \textbf{Model Type} & \textbf{Classification Performance} & temp & \textbf{Compression Performance} & temp & temp \\
    temp & \acs*{MAP}\TblrNote{1} Score & Beam Prediction Accuracy & The Number of Parameters & Model Size & \acs{FPS}\TblrNote{2} \\
    No-Pruning Model & 85.12 & 90.34\% & 61.52~\myunit{M} & 236.52~\myunit{MB} & 39 \\
    Channel Pruning Only Model & 84.68 & 90.15\% & 12.64~\myunit{M} & 48.32~\myunit{MB} & 77 \\
    Channel and Layer Pruning Model & 84.42 & 90.06\% & 11.17~\myunit{M} & 42.71~\myunit{MB} & 91
\end{talltblr}
\end{table*}

\subsubsection{Dataset Validation from Visual and Wireless Aspects}

In the proposed framework of vision-assisted beam management, object detection plays a crucial role in both the existence detection and beam selection tasks. Thus, Fig.~\subref*{fig_sub2} presents the validation results for datasets from the visual perspective. Explicitly, it shows a single image frame labelled by the bounding boxes of three moving \acp{UE} in cars. The accuracy of the labeling results demonstrates that the visual datasets generated accurately characterise the movement of objects in each frame, and that the proposed vision-assisted beam management framework can also effectively track objects in real time.

On the other hand, Fig.~\subref*{fig_sub3} illustrates the signal power of the randomly generated rays between the \ac{BS} and vehicular \acp{UE}, which comes from the wireless datasets. The larger the distance between rays, the lower the received power, which confirms the trends of the generated wireless datasets. Additionally, due to the mobility of various objects, wireless datasets can occasionally contain zero data. For example, whenever the car farthest from the \ac{BS} runs into the shadow of a bus, no rays are detected for this frame, resulting in beam tracking outage.

\subsubsection{Results for Inadequately Labeled Data}

Fig.~\subref*{fig_sub4} shows the beam prediction accuracy of the improved models using metric learning in the case of inadequately labeled data. For each category, the metric learning normally requires only one feature template. Nevertheless, there may be $ K $ samples in each category, thereby having $ K $ feature vectors. Therefore, the $ K $ feature vectors should be combined to produce a representative category template. Three schemes are considered for comparison, i.e., Version 3 of \ac{YOLO}, the average metric learning, and the $ k $-means-weighted average metric learning. Specifically, the average metric learning combines all feature vectors using the arithmetic mean method over $ K $ samples. To overcome the homogenization of arithmetic mean,  the $ k $-means-weighted average metric learning is also studied, in which $ K $ samples are first classified by the $ k $-means method, and the cluster features obtained are then combined by the weighted averaging method.

It is clearly shown that both metric learning schemes perform better than \ac{YOLO}v3, achieving an accuracy of about $ 84.12\% $ with only 10-shot learning. Furthermore, the $ k $-means-weighted average metric learning slightly outperforms the average metric learning. In conclusion, the metric learning schemes are more efficient than \ac{YOLO}v3  when data are inadequately labelled.

\subsubsection{Results for Lightweight Compression}

The performances of the improved \ac{ML}-based models having different lightweight compression are also evaluated. Prior to discussing the simulation results, we briefly highlight the performance metric, i.e., \ac{MAP} score. As a derivative of the \ac{AP}, \ac{MAP} is the average of \ac{AP} score, while the \ac{AP} score generally is obtained by calculating the area under the \ac{PR} curve. To summarize, the \ac{AP} score is calculated for each category, then averaged to determine the final \ac{MAP} score.

Table~\ref{table_prune-perform-comp} presents the simulation results for the cases of no-pruning, channel pruning only, as well as channel-pruning and layer-pruning. As illustrated in Table~\ref{table_prune-perform-comp}, the classification performance of both pruning models degrades compared to the no-pruning model. However, the accuracy erosion is modest for two pruning models. For instance, with regard to the channel and layer pruning model, the \ac{MAP} performance and beam prediction accuracy only deteriorates by about $ 0.8\% $ and $ 0.3\% $, respectively. By contrast, the pruning operation results in a significant model size reduction and an acceleration of the inference speed. The number of parameters and the model size are reduced by about $ 82\% $, which is more beneficial for the practical deployment of latency-sensitive applications.

\section{Open Discussion}

\subsection{Combination with Hierarchical Beam Search}

Usually, the explicit training required for finding the best beam directions in the angular domain is indispensable. In contrast to the classical exhaustive search based training, hierarchical training has been proposed as a promising technique of reducing both the complexity and the overhead. However, a trade-off must be struck between the phase shift resolution of training and the complexity imposed. For example, when a low phase shift is chosen for the first stage of training, the beam direction can be selected more accurately. However, this imposes higher feedback delay and higher overhead, or vice versa. To strike a compelling trade-off, a vision-assisted beam management scheme can be used as the first stage of training, because it does not rely on \ac{UE} feedback for beam selection. Subsequently, the accuracy of beam search can be further improved through a fine-tuning of \ac{CSI}-based beam management along with a lower phase shift in the following training stage.

\subsection{Uplink and Downlink Beam Matching}

As a result of the propagation differences between the uplink and downlink, especially for \ac{FDD} systems, the downlink beam selection based on the uplink channel estimation operation usually requires calibration to improve accuracy. On the other hand, the location of the user can be accurately determined by vision-assisted beam management regardless of the frequency band. Therefore, how to use this information to support beam matching on both the uplink and downlink becomes a very interesting topic.

\subsection{Dual-Band Communications with Sub-6 GHz}

Recently, the dual-band communication mode including \ac{mmWave} and sub-$ 6~\myunit{GHz} $ communications is becoming increasingly popular. Therefore, another open challenge is how to exploit the extra information at sub-$ 6~\myunit{GHz} $ so as to enhance the \ac{mmWave} beamforming performance. Intuitively, the proposed vision-assisted \ac{mmWave} communications depends on having \ac{LOS} propagation for its accurate operation, and it is vulnerable to blockage. For instance, when multiple \acp{UE} are captured by the camera without any additional details, vision-assisted beamforming may falter. On the other hand, sub-$ 6~\myunit{GHz} $ communications generally works well for both \ac{NLOS} and \ac{LOS} channels, and it is capable of providing the related control information, including \ac{CSI} and other user-specific information. This information can assist in the detection of active \acp{UE} and multi-user discrimination when using vision-assisted \ac{mmWave} communications. Additionally, for further reducing the complexity of exploiting sub-$ 6~\myunit{GHz} $ communications, the above-mentioned hierarchical beam search technique can be used for sub-$ 6~\myunit{GHz} $ to provide prompt user-specific information.

\subsection{Multi-Cell Beam Management}

The coverage distance of \ac{mmWave} communications is typically small, and \acp{UE} often appear at the cell edge. The beam selection of cell-edge \acp{UE} can be handled more accurately by adopting vision-assisted beam management. Specifically, the videos obtained by the cameras of multiple adjacent \acp{BS} can be processed jointly. Due to the fact that the same \ac{UE} is captured in multiple images at the same time, its position can be more accurately determined using \ac{ML} algorithms as well as the beam direction. By aligning the beams of two adjacent \acp{BS} for the target \ac{UE}, a more reliable communication connection can be achieved. The issues associated with channel feedback overhead can be avoided by such a vision-aided multi-cell beam management.

\section{Conclusions}
\label{sec:Conclusion}

Vision-assisted beam management is paving the way for improved \ac{mmWave} communications by relying on machine learning models of analyzing visual data. This enables us to tackle several important challenges of \ac{ML}-based model implementation for \ac{mmWave} beamforming. In particular, sophisticated network pruning has been used to compress the models for reducing the complexity. Additionally, a model based on metric learning has been shown to be an effective option for dealing with the problem of inadequately labeled data in practical applications. A \ac{ML} model based on image sequences has also been conceived for multi-user scenarios and to mitigate the blockage problems. Then, an animation modeling software and ray-tracing software were used for successfully building a new simulation platform to generate various labeled visual and wireless data for performance evaluation and model validation. Our simulation results show that \ac{ML}-based models work well with vision-assisted \ac{mmWave} beam management schemes. Furthermore, some open challenges are presented for the guide of future research works. Additionally, \ac{DFRC} systems may be capable of simultaneously performing wireless communications and remote sensing, when they become available but have a huge complexity. The alluring topic of combining these technologies is also interesting for future research. Suffice to say that the vision-based system investigated in this treatise only requires a low-cost camera and object-recognition software.

%
\bibliographystyle{IEEEtran}
\bibliography{IEEEabrv,Bib/Ref}
%
%
\end{document}